# Incubation Induced Light Concentration Beyond the Diffraction Limit for High-Resolution Glass Printing

Haisu Zhang[1, ‡, *], Peng Wang[2, 3, 4, ‡], Wei Chu[1, *], Jianping Yu[3, 4], Wenbo Li[3, 6], Jia Qi[3,4], Zhanshan Wang[2] and Ya Cheng[1, 3, 7, *]

[1]State Key Laboratory of Precision Spectroscopy, East China Normal University, Shanghai 200062, China
[2]School of Physics Science and Engineering, Tongji University, Shanghai 200092, China
[3]XXL-The Extreme Optoelectromechanics Laboratory, School of Physics and Electronic Sciences, East China Normal University, Shanghai 200241, China
[4]State Key Laboratory of High Field Laser Physics, Shanghai Institute of Optics and Fine Mechanics, Chinese Academy of Sciences, Shanghai 201800, China
[5]University of Chinese Academy of Sciences, Beijing 100049, China
[6]School of Physical Science and Technology, ShanghaiTech University, Shanghai 200031, China
[7]Collaborative Innovation Center of Extreme Optics, Shanxi University, Taiyuan, 030006, China
‡These authors contributed equally to this work.
*Correspondance: hszhang@phy.ecnu.edu.cn, wchu@phy.ecnu.edu.cn, ya.cheng@siom.ac.cn.

**In the past two decades, tremendous efforts have been exerted to understand and control the delivery of ultrashort laser pulses into various types of transparent materials ranging from glass and crystal to polymer and even bio-materials. This approach opens up the route toward determinative and highly localized modification within the transparent materials, enabling three-dimensional (3D) micromachining of the materials into sophisticated structures and devices with the extreme geometrical flexibility. Owing to the linear diffraction and nonlinear self-focusing effects, the focal volume typically exhibits an asymmetric profile stretching along the longitudinal direction. This effect becomes more severe when focusing deeply into the transparent substrates for printing objects of large heights. In this work a new laser-material interaction regime is identified with the exceptional incubation effect originating from self-regulated multiple-pulse interactions with accumulated material changes. Our finding reveals a focal-volume-invariant modification deeply inside the fused silica glass, in striking contrary to the traditional believes that the geometrical shape of the laser induced modification follows the intensity distribution of the inscription laser. A macro-scale geometrically complex glass sculpture is successfully manufactured with the incubation assisted ultrashort laser inscription at uniform micrometer resolutions in all three dimensions.**

Ultrashort laser inscription is the ideal workhorse for high-resolution three-dimensional material machining[1,2], covering a broad spectrum of application areas ranging from micro-optoelectromechanics integration and biomimetic device fabrication to large-capacity data storage and quantum information science[3-6]. The ultrashort pulse duration minimizes thermal effects accompanying laser-material interactions and the corresponding ultrahigh peak intensity facilitates efficient energy deposition by free carrier absorption irrespective of the bandgap of the material under processing[7]. Since the initial free carriers are generated by the extremely nonlinear photo-ionization process effectively confined within the focal volume, the localized energy deposition by tightly focused ultrashort laser pulses features a micrometer-to-nanometer processing resolution roughly determined by the geometrical size of the laser focus[8, 9].

Tremendous efforts have been put to control the delivery of light into bulk materials and optimize the focus size to approach and even surpass the ideal spatial resolution of laser inscription set by the diffraction limit of light. Both active control schemes including slit-shaping[10], double-pulse excitation[11], simultaneous spatiotemporal focusing[12], and passive optimization loops by adaptive wavefront corrections with spatial light modulators[13, 14], are demonstrated to maintain the desired focal profiles and processing resolutions within few millimeters' depth inside bulk materials. However, in the scenario of fabricating centimeter-scale-height macro-objects utilizing a low numerical aperture (NA) lens of a long working distance, all the methods mentioned above degrade rapidly. Since the transversal and longitudinal sizes of the focus are respectively proportional to $NA^{-1}$ and $NA^{-2}$ of the focal lens in the linear interaction regime[15], the use of low-NA lens largely increases the attainable focal size especially for the longitudinal dimension. The spherical aberration induced by the index-mismatch at the interface and the nonlinear propagation effects at high input powers necessitated by the low-NA focusing[16, 17], can further deteriorates the focal profile which can only be suppressed by pulse shaping to a limited extent.

In this work, we present pump-probe investigations of ultrashort laser inscription in fused silica glass. By carefully comparing the transient dynamics launched by femtosecond and picosecond laser pulses, clear evidence of multiple-pulse incubation effect facilitating highly localized energy deposition decoupled from the laser focal profile is revealed for the picosecond laser inscription. Specifically, the achieved longitudinal resolution of picosecond laser inscription is 5 times smaller compared to the confocal range of the focusing lens. An exquisite sculpture with few-centimeter height, reproducing the Chinese treasure *Four-sheep statue*, is successfully fabricated at ~20 μm resolutions in all three dimensions. The discovered ultrashort laser-material interaction in the picosecond regime features synergetic actions of multiple pulses with accumulated material changes such as electronic defects, structural rearrangement and mechanical damages, on the spatial localization of light. Importantly, the localized energy deposition and material modification by picosecond laser pulses is self-regulated and thus the achieved resolution is robust to input energies fairly above the optical breakdown threshold, in sharp contrast to the sub-diffraction-limit femtosecond laser inscription uniquely occurring at near-threshold energies[8]. As a consequence, picosecond laser inscription enables unparalleled high-throughput rapid-prototyping of large objects with industry-compatible fabrication scalability.

## Results and discussions

### Pulse duration dependence of ultrashort laser inscription

The optical transmission micrographs from the cross-sectional view of the laser inscribed straight lines inside fused silica are shown in the front side of Fig. 1, where the pulse duration of the inscription laser is increased from the minimum value (annotated as '0') by adding opposite frequency chirps. The transverse inscription scheme is employed where the glass sample is moving horizontally across the laser focus at the constant velocity of 10 mm/s. It is clearly seen that nearly spherical cross-sections of ~20 μm diameter for the laser-inscribed lines are obtained as long as the pulse duration of the inscription laser is longer than 5 ps. Faintly visible longitudinally elongated modification is obtained by the femtosecond laser inscription, in contrary to the high-contrast modifications induced by the picosecond laser inscription at the same laser pulse energy (~10 μJ) and focusing condition (NA=0.10).

The simulated intensity distribution of the inscription laser focused 10 mm underneath the glass surface (linear propagation is supposed) is depicted in the inset of Fig. 1 for comparison[18], where the laser focus is characterized by its longitudinal and transversal sizes separately given by twice the Rayleigh range $\Delta Z = 2Z_R = 2n_0\lambda/\pi(\text{NA})^2 = 95\ \mu m$ and twice the focal radius (defined at the $1/e^2$ intensity) $2\omega_0 = 2\lambda/\pi(\text{NA}) = 6.6\ \mu m$[15]. Here $n_0$ is the refractive index of the glass at the wavelength $\lambda$ of the inscription laser. It is clearly demonstrated that the spatial distribution of material modifications induced by picosecond laser pulses with sufficient durations decouples from the focal profile of the inscription laser, i.e. the modified region is highly confined within one-fifth of the longitudinal extension of the laser focus, while the transversal dimension of material modification is three times larger than the focal diameter. Apparently, this surprising behavior cannot be simply ascribed to the thresholding effect due to the highly nonlinear nature of the laser inscription process, whose direct consequences are material modifications beyond the diffraction-limits in both the longitudinal and transversal dimensions.

### Transverse pump-probe microscopy

To investigate the underlying mechanism of this peculiar laser inscription regime, transverse pump-probe microscopy is employed to visualize the spatiotemporal plasma dynamics and structural transformations in laser-excited glass with both femtosecond and picosecond pulses. The detailed experimental setup is introduced in the method section. Briefly, 370 fs laser pulses at 1030 nm wavelength are split into the pump and the probe. The pump pulse is loosely focused (NA=0.10) at ~10 mm depth inside the fused silica glass after passing the double-pass grating stretcher for pulse-duration manipulation, while the probe pulse is mechanically delayed and frequency doubled to 515 nm and then used as the strobing light to illuminate the pump interaction volume from the transverse direction (as shown in Fig. 2(a)).

Single-pulse interaction regime

Transient transmission images captured by single-shot probe pulses at different delays with respect to the pump are shown for the pump pulse duration of 370 fs in Fig. 2(b) and 10 ps in Fig. 2(c), respectively. Pump pulses incident from the left are set at the same energy (20 μJ) for both durations. Each image is normalized with the reference image taken before the pump incidence at pristine positions. Different plasma generation sequences can be clearly noticed by comparing the transmission images taken at short delays shown in Fig. 2(b) and Fig. 2(c), where free electrons generated by the pump pulse induce remarkable absorption of the traversing probe pulse. Plasma absorption commences around the geometrical focus at the leading edge of the 10 ps pump pulse and extends towards the pump incident direction following the peak and trailing edge of the pulse, whereas the prefocal plasma generation with subsequent forward growth along the pump incident direction is established by the 370 fs pump pulse. This is in well consistence with the moving-breakdown model predicting opposite optical breakdown sequences for femtosecond and picosecond laser pulses[19].

Free electrons injected into the conduction band of fused silica are quickly trapped in ~150 fs by the deformed lattice potential due to the scission of Si-O bonds[20], forming immobile electron-hole pairs localized at specific lattice sites-the self-trapped excitons (STE)[21]. The observed rapid recovery of probe transmission at 2 ps in Fig. 2(b) reflects the fast electron trapping process after the passage of the 370 fs pump pulse. The STEs further relax in a sub-picosecond timescale into a variety of point defects such as the well-known nonbridging Oxygen hole centers (NBOHC) and the Oxygen vacancy centers (E') in fused silica[22, 23]. Defect accumulation in turn induces large displacement and localized heating of the lattice, leading to the build-up of thermoelastic stress in a few picoseconds[24]. The strong revival of transmission loss starting from ~4.6 ps in Fig. 2(b) can be ascribed to the thermal stress induced light scattering as well as defect absorption. The localized stress is then gradually released by structural rearrangement and pressure-wave formation in few hundreds of ps to a few ns, which is manifested as the partial recovery of transmission at 230 ps and 330 ps in Fig. 2(b).

On the contrary, the lack of rapid electron trapping during and after the 10 ps pump pulse as revealed by the lack of obvious transmission recovery shown in Fig. 2(c), indicates that the pump laser generates a higher density of free electrons, rendering the lattice severely distorted and deformed which is unable to trap electrons. It is well accepted that picosecond pulses with adequate intensities can provide sufficient acceleration of free electrons generated by direct photo-ionizations through the inverse bremsstrahlung[25]. The avalanche ionization emerges for the high-energy portion of heated electrons with excess energy than the bandgap. The combined action of both ionization processes induces continuous increase of free electron density as well as the collisional heating of both electrons and lattice during the pump pulse. The irradiated part of glass after the pump pulse features strongly delocalized molecular bonds and probably the onset of a low viscosity state with vibrational activity due to rapidly elevated temperature[26], which are supported by the long-lived plasma absorption lasting hundreds of ps shown in Fig. 2(c).

The observed plasma absorption and structural transformation regions induced by single pump pulses shown in Figs. 2(b) and 2(c) basically resemble the highly elliptical laser focal profile, with subtle differences arising from the spatial energy depositions and nonlinear propagation effects due to the competition of self-focusing and plasma defocusing setting in at high laser powers[17,27]. The critical power ($P_{cr}$) for self-focusing in fused silica at the pump wavelength (1030 nm) is given by $P_{cr} = 3.77\lambda^2/8\pi n_0 n_2$[17], which equals to ~4 MW using the linear and nonlinear refractive index $n_0 = 1.45$ and $n_2 = 2.7 \times 10^{-20}\ m^2/W$ [28]. The used peak powers of the pump pulses are $0.5 P_{cr}$ and $13.5 P_{cr}$ for the 10 ps and 370 fs pulses respectively. Substantial nonlinear propagation effects are thus expected to take place in the femtosecond laser-glass interaction, as evidenced by the extended lengths of the transient plasma and structural transformation shown in Fig. 2(b). No localization of light is observed in the context of single-pulse interaction for both pulse durations, which results in hardly observable material modifications induced by single pump pulses due to insufficient energy depositions within the broad focal volume. Multiple-pulse irradiation is needed to generate remarkable modifications as shown in Fig. 1, underlining the significant incubation effect of multiple-pulse interactions in the laser inscription process.

### Multiple-pulse incubation regime

Further pump-probe microscopy is concentrated on the incubation effect of laser-glass interaction. To this end, each glass site is pre-irradiated by certain number of pump pulses, after which the transient plasma and structural dynamics induced by the succeeding pump pulse is captured by single-shot probe pulses at various delays. The transient images taken at the delays of 2 ps and 10 ps, representing the plasma dynamics during the 10 ps pump pulse, are shown in Figs. 3(a) and 3(c) respectively. Distinct features are unfolded as the spatial shift and confinement of plasma absorption towards the incident direction following accumulation of pump pulses. It implies that the propagation of subsequent pulses through the region modified by previous pulses is arrested for the 10 ps pump pulse. The permanent modifications generated by seven pulses are clearly observed and depicted as gray-scale images at the bottom of Figs. 3(a) and 3(c). Severe material disruptions are present at regions corresponding to the localized plasma absorption in the transient images.

Similar measurements are also conducted for the 370 fs pump pulse at the probe delays of 800 fs and 10 ps, respectively, corresponding to the moments of maximum plasma generation and build-up of thermoelastic stress as revealed in Fig. 2(b). The results are shown in Figs. 3(b) and 3(d), featuring similar spatial extensions of plasma and thermal stress generated by each single pulse irrespective of previous exposures. The apparent incapability of energy concentration by irradiation with multiple femtosecond pulses causes the final material modification after the 7$^{th}$ pulse still weakly detectable as shown at the bottom of Figs. 3(b) and 3(d).

It is clearly unveiled that the incubation effect plays an essential role in laser-glass interactions with picosecond pulses, in striking contrast to its counterpart in femtosecond pulse interactions. The electron binding energy in the irradiated part of glass is reduced compared to the pristine position[29], due to the presence of defect centers, structural rearrangement as well as residual damage after laser exposure. Photoionization in the modified region thus emerges at a lower intensity threshold, facilitating quick generation of free electrons in the leading edge of excitation pulse. This in turn spares more times for electron heating by the remaining pulse and thus efficiently promotes ensuing avalanche ionization to approach the critical electron density and deposit most of incident laser energy by the picosecond pulse rather than the femtosecond pulse. This hypothesis is further supported by the observation of bright plasma continuum emissions during bremsstrahlung emanated from the localized material modification area induced by multiple picosecond pulses instead of femtosecond pulses[30]. Consequently, highly selective spatial energy depositions are established by accumulated positive feedbacks from material changes during the multiple pulse interactions in the picosecond regime. The incubation induced light and energy localization exists close to the deeply modified region by preceding pulses, and the energy transport beyond this region is suppressed.

## Incubation assisted picosecond laser inscription beyond the diffraction-limit

To further test the paradigm of incubation induced light concentration, the material modifications generated at variable exposures of picosecond pulses are investigated. The transverse transmission micrographs of the permanent modifications of glass, induced by sequential irradiation of 500 pulses with the pulse duration of 10 ps, the single pulse energy of 10 μJ and the repetition rate of 25 kHz, are shown in panels (i)-(viii) of Fig. 4(a). The stable material modification after irradiation of 30,000 pulses is shown in panel (ix) for comparison. It is clearly seen that the optical breakdown region is shifted far from the laser focus following the continuous exposures of laser pulses, and the material modification grows longitudinally towards the laser incident direction until reaching the intensity region below the breakdown threshold. Especially, the material modifications induced by each exposure of pump pulses are highly localized near the previous modified regions, i.e. one can compare the orange arrows in each panel which mark the delimiter of the newly-created material modification. It is thus confirmed that a sequential collapse of pump pulses towards the incident direction is responsible for the observed elongated stable material modification shown in panel (ix).

The observed material response to the varying exposure unambiguously signifies the critical multi-pulse incubation effect on the picosecond laser inscription. The self-regulated growth of material modifications can be also achieved when the glass sample is continuously scanned across the laser focus. It is clearly demonstrated that the longitudinal growth of material modifications towards the inscription laser at the fixed site of glass (Fig. 4(a)), is turned transversely towards the sample moving direction, as marked by yellow dotted lines in Fig. 4(b) when the sample is moved from bottom to up and in Fig. 4(c) when the sample is moved from up to bottom. Material modifications induced during sample movements are well confined within a spherical cross-section of ~20 μm diameter inside the glass as shown in the insets of Fig. 4(b) and Fig. 4(c). In particular, these sub-diffraction-limit bulk modifications persist as long as the sample is moving at a wide range of velocities depending on the specific pulse duration of the inscription laser[30].

A macroscopic sculpture reproducing the Chinese treasure *Four-sheep statue*, is fabricated by the incubation induced high-resolution laser printing in a 4 cm-height fused silica glass, assisted by post-processing mechanical polishing and chemical wet etching process[31]. The full fabrication procedure is illustrated in Fig. 5(a) with the detail described in the Method section. The digital model of the sculpture is depicted in Fig. 5(b), and the fabricated sculpture is shown in Fig. 5(c) and 5(d) from the inclined-view and the front-view, respectively. The zoomed-in details are further demonstrated for (e) the attached decorative pattern, (f) the lateral protrusion, (g-h) the horns of the front dragon and the corner sheep, respectively. The high-fidelity duplicate of the exquisite Chinese treasure proves the decent capability of the established 3D large-scale glass-printing technique enabled by high-throughput picosecond laser inscription.

Superiorities of picosecond laser inscription over femtosecond laser inscription are well recognized before. Theoretical works have pointed out that in comparison to femtosecond pulses, picosecond pulses at the same energy can deliver higher fluences into the confocal region due to reduced prefocal energy depletions as well as shortened plasma defocusing events which increase the effective irradiation of material around the focus[32, 33]. Meanwhile, few-picosecond pulses compete well with femtosecond pulses in the context of minimum collateral damages during the inscription process, considering the sluggish thermomechanical responses of irradiated materials in few hundreds of ps to a few ns. Recent works demonstrate that picosecond pulses at mid-infrared wavelengths deeply focused inside silicon wafer enable high-precision bulk modifications, holding great promises to current electronic industry[34, 35].

## Conclusion

In this work a new laser-glass interaction regime is identified as the exceptional multiple-pulse incubation effect for picosecond laser inscription in fused silica glass. Transverse pump-probe microscopy is employed to reveal the plasma and structural dynamics deep inside the glass excited by loosely focused laser pulses of different durations, supporting the incubation assisted energy concentration of few-picosecond pulses beyond the diffraction limit of the focusing lens. This observation brings an unprecedented method to print macroscopic glass samples at microscopic resolutions with arbitrary geometrical complexities, and is doomed to accelerate the industrial applications of laser micromachining to broad perspectives of high-throughput massive manufacturing.

# Methods

**Experimental set-up.** High-power femtosecond pulses from a fiber laser-amplifier system (FemtoYL™), with the central wavelength of 1030 nm, Fourier-transform-limited duration of 370 fs (full width half maximum in intensity), and repetition rate of 25 kHz, are first split into the pump pulse and the probe pulse. The pump pulse can be temporally broadened with a home-built double-pass grating stretcher (two gold-coated gratings with 1200 line/mm). The probe pulse is first frequency converted into 515 nm by a 400 μm thickness type-I BBO crystal. The pump pulse is focused deeply inside the fused silica glass by a NA=0.10, 5× microscope objective (MPlan N, Olympus), while the collimated probe beam illuminates the glass from the transverse direction. Mechanical delay line with motorized stage is inserted to control the delay of the probe pulse with respect to the pump pulse. A NA=0.40, 20× microscope objective (MPlanFL N, Olympus) is used to image the transmitted probe photons with the help of a field-lens of 15 cm focal-length onto a high-resolution CMOS camera (Lw115C, Lumenera). A bandpass filter (FL514.5-3, Thorlabs) centered at 514.5 nm is inserted before the field-lens to block the diffused pump photons as well as the photo-luminescence from the interaction region. Alternatively, a short-pass filter (FES0800, Thorlabs) is used when measuring the photo-luminescence induced by the pump pulse.

**Data acquisition.** Single-shot laser pulses selected by the built-in pulse-picker are used for the transverse pump-probe microscopy. Reference images $I_0(x,y)$ are first taken by a single probe pulse at each pristine position of the glass, and then the same positions are simultaneously illuminated by a pair of pump and probe pulses with various delays to capture the transient images $I_t(x,y)$. For the single-pulse interaction regime, the glass is moved to fresh sites after each exposure of pump pulses to avoid accumulation effects. To measure the multiple-pulse incubation effect, the glass is kept at the same position with incremental exposure of pump pulses. Each transient image is normalized with corresponding reference image and the final transmission images shown in the main text are colored by the transmission rate of probe pulse expressed in $10\log_{10}(I_t/I_0)$ [dB].

**Macro-object fabrication.** Electronic 3D-model of *Four-sheep statue* is first embedded and subtracted in a cuboid frame. The inverse-converted model is sliced into horizontal planes with a fixed slice thickness of 50 μm. Then a rapid raster-scan of the laser focus, along the pre-designed paths layer by layer from the bottom to the top of the glass, is conducted by the high resolution 3D motion stages at the velocity of 50 mm/s (ABL15020 for horizontal motion, ANT130-110-L-ZS for vertical motion, Aerotech Inc.). Afterwards the glass sample is polished to remove the outer part of irradiated areas, and then the polished sample is immersed in a wet-etching bath of potassium hydroxide (KOH) with a concentration of 10 mol/L at a temperature of 90 °C for tens of hours, to selectively remove the laser irradiated part of glass. The picosecond pulses at the repetition rate of 200 kHz, the pulse duration of 10 ps, and the single-pulse energy of 8 μJ, are focused by the long working distance (37.5 mm) objective with NA=0.14, 5× (M Plan Apo NIR, Mitutoyo Corporation).

## Acknowledgements

This research was funded by the Key Project of the Shanghai Science and Technology Committee (Grant Nos. 18DZ1112700, 17JC1400400), National Natural Science Foundation of China (Grant

61590934, 11734009, 11674340, 11822410), National Key R&D Program of China (Grant No. 2019YFA0705000) and the Strategic Priority Research Program of Chinese Academy of Sciences (Grant No. XDB16030300), and Key Research Program of Frontier Sciences, Chinese Academy of Sciences (Grant No. QYZDJ-SSW-SLH010).

## Author contributions

W. C. and Y. C designed the project. H. Z built the pump-probe microscopy setup. H. Z and P. W performed the pump-probe experiment. H. Z. analyzed the data. P. W and W. C fabricated the macro-scale sculpture. H. Z and Y. C wrote the first manuscript. Z. W and Y. C supervised the project. All authors discussed the results and contributed to the final manuscript.

## Competing interests

The authors declare no competing interests.

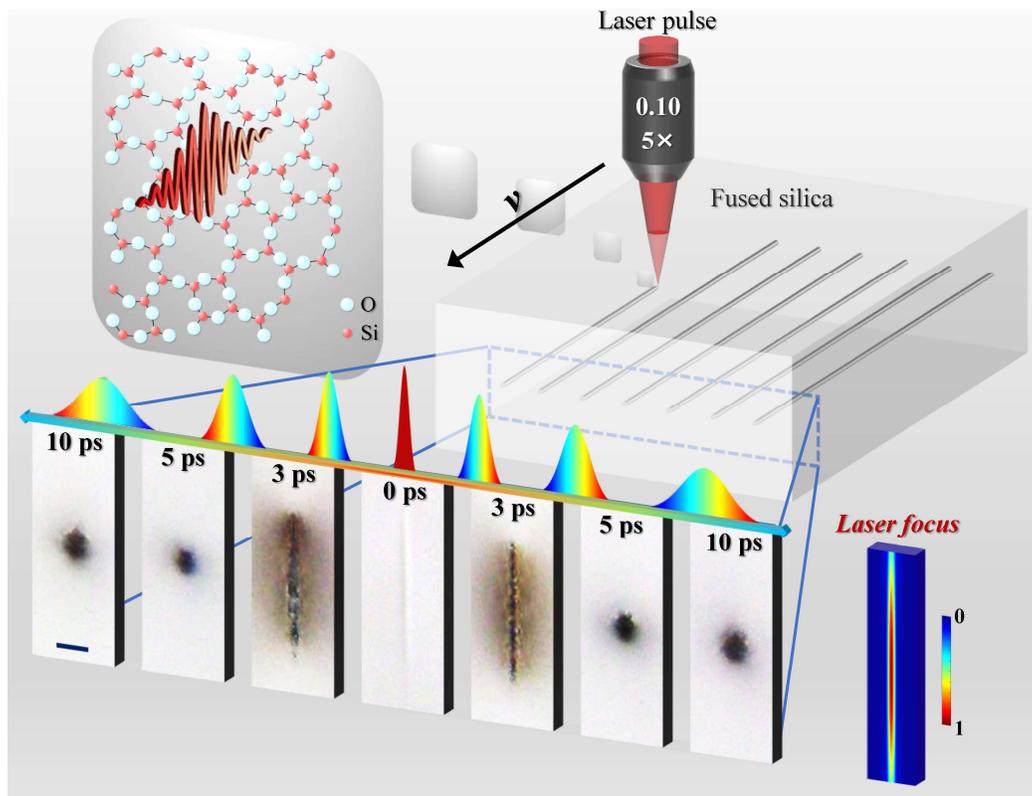

**Figure 1**. Cross-sectional transmission micrographs of the laser-inscribed straight lines in fused silica glass. The pulse duration of the inscription laser is annotated on the top part of each micrograph. The simulated laser focal profile is shown in the right inset. The scale bar is 20 μm and applies for all the micrographs and the laser focus.

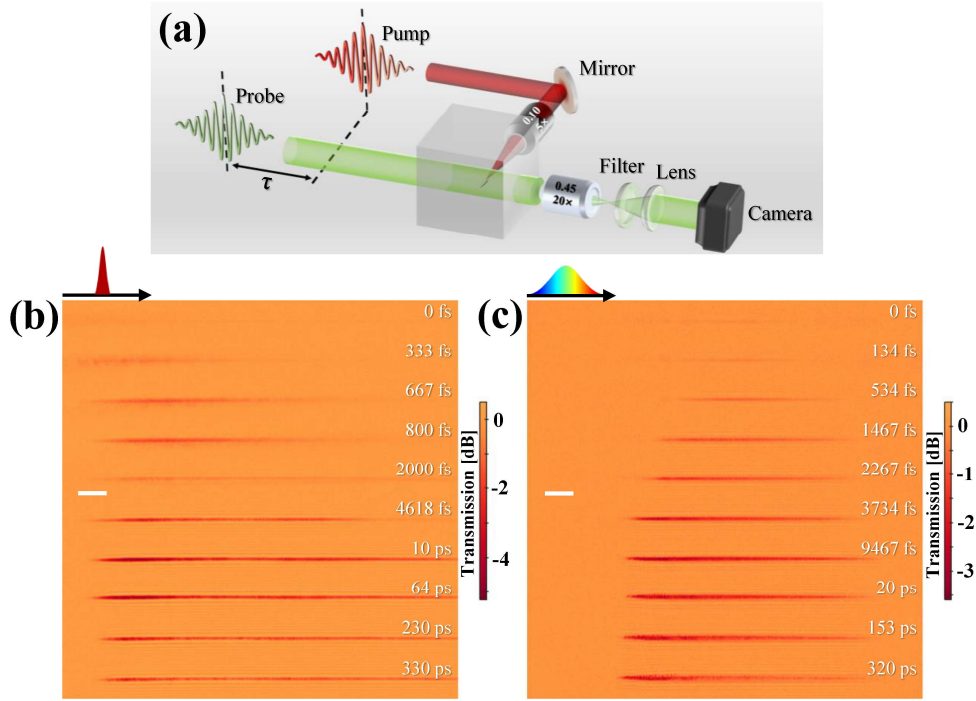

**Figure 2**. (a) Schematic of the transverse pump-probe microscopy. (b-c) Transverse pump-probe microscopy in the single-pulse interaction regime. The single-shot probe transmission images captured for the 10-ps pump pulse (b) and the 370-fs pump pulse (c). The delay of the probe pulse with respect to the pump pulse is labelled in (b) and (c). The scale bars in (b) and (c) are 20 μm.

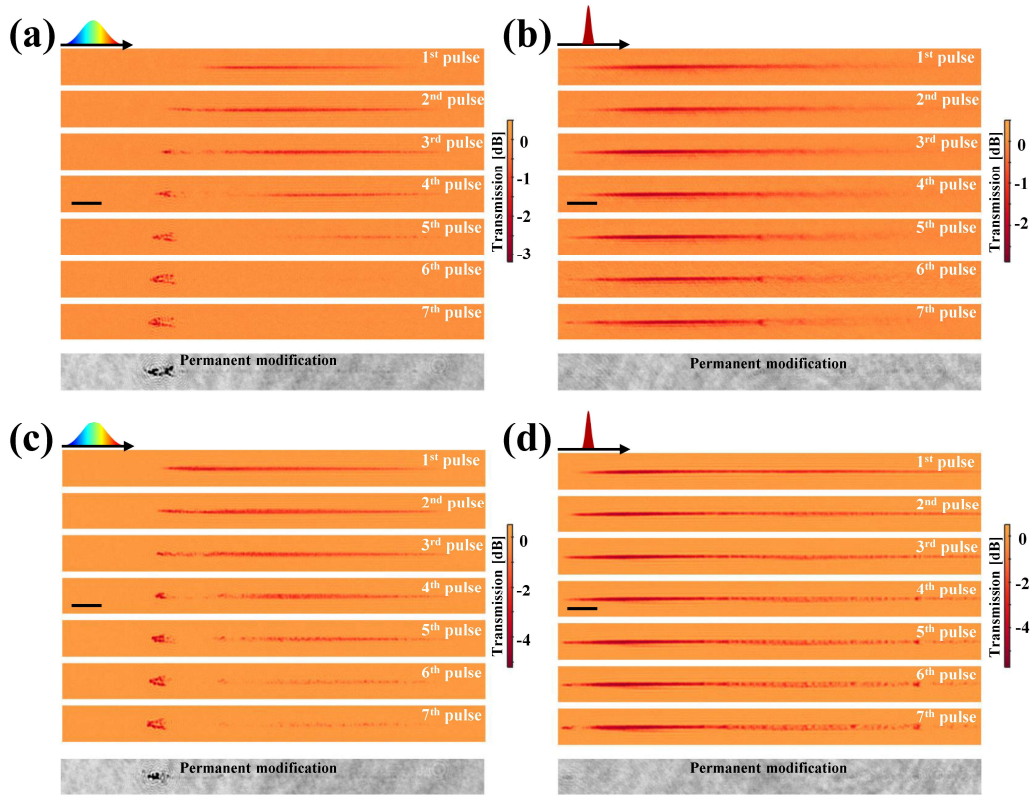

**Figure 3**. Transverse pump-probe microscopy in the multiple-pulse interaction regime. The single-shot probe transmission images captured for the 10-ps pump pulse (a, c) and for the 370-fs pump pulse (b, d). The delay of probe pulse with respect to the pump pulse is 2.2 ps in (a), 0.8 ps in (b), 10 ps in (c) and (d). The corresponding permanent material modifications are shown in the bottom of each figure. The scale bars in each figure are 20 μm.

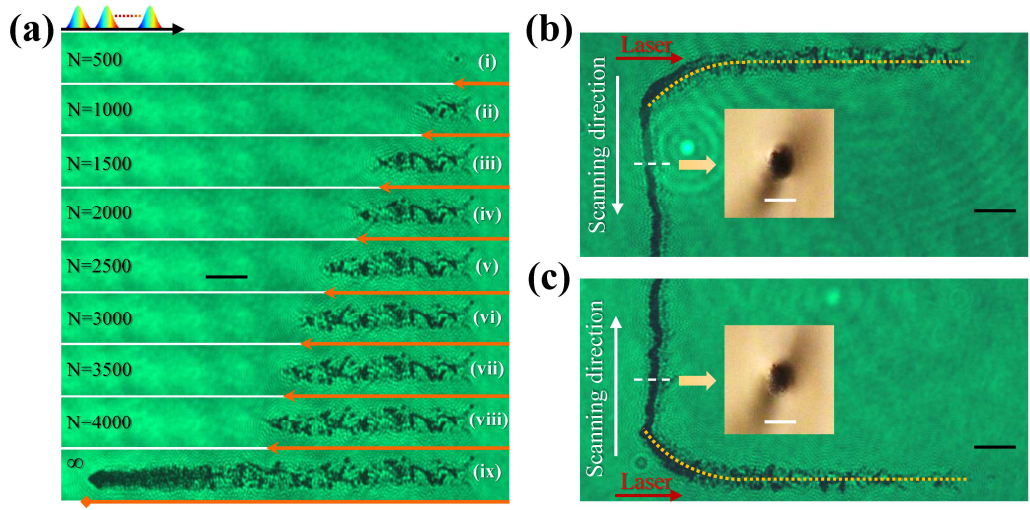

**Figure 4**. Transmission micrographs of permanent material modifications induced by picosecond laser inscription. (a) The permanent material modifications after sequential exposures of pump pulses on the same glass position (the exposed pulse numbers increase from (i) to (viii), and the stable permanent modification is shown in (ix)). (b-c) The material modifications induced by continuously moving the glass sample across the laser focus when the sample is moved from bottom to up (b) and the sample is moved from up to bottom (c). The cross-sectional micrographs are shown in the insets of (b) and (c). The scale bars in each figure are 20 μm.

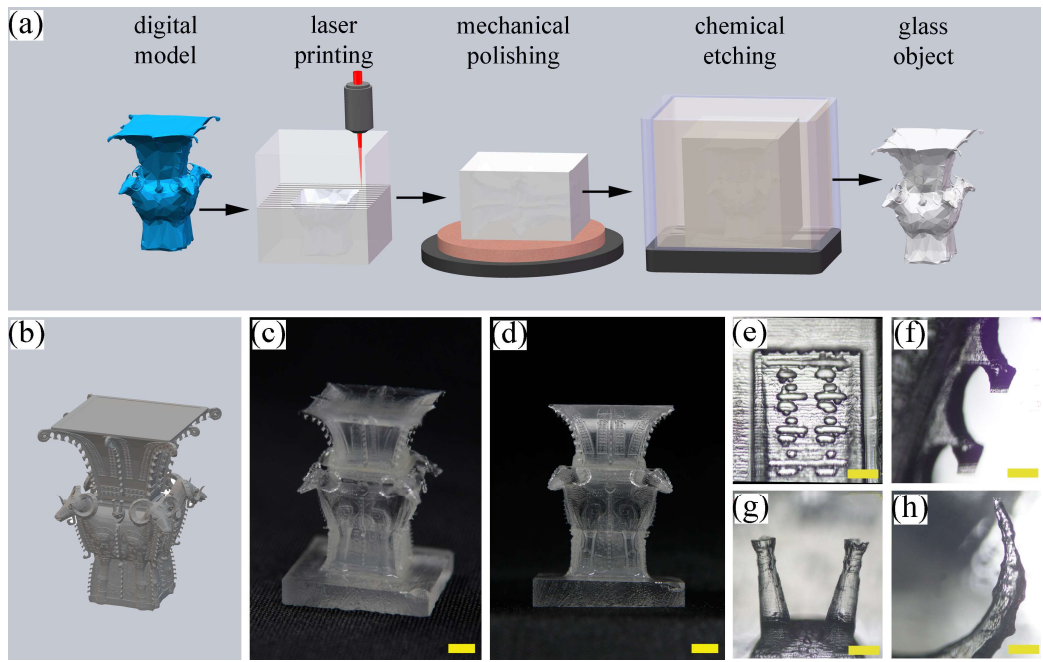

**Figure 5.** Large-scale glass-object printing by picosecond laser inscription assisted wet etching. (a) The fabrication procedure. (b) The model of *Four-sheep statue*. (c-d) The inclined-view and the front-view of the fabricated sculpture. (e-h) The zoom-in details of the fabricated sculpture. The scale bars in (c-d) are 5 mm, and the scale bars in (e-h) are 500 μm.

16